\documentclass[twocolumn,prd,showpacs,10pt]{revtex4}
\usepackage{graphicx}
\usepackage{epsfig}
\usepackage{symb} 

\def\be{\begin{equation}}
\def\ee{\end{equation}}
\def\bea{\begin{eqnarray}}
\def\eea{\end{eqnarray}}
\def\LL{\mathcal{L}}
\def\pp{\theta}
\def\dat{D}
\def\pop{\tau}
\def\Ia{\mathrm{Ia}}
\def\Ibc{\mathrm{Ibc}}
\def\II{\mathrm{II}}

\def\eff{\mathrm{eff}}
\def\hpm{\hphantom{-}}
\begin{document}

\title{Bayesian Estimation Applied to Multiple Species: \\Towards cosmology with a million supernovae}


\author{Martin Kunz}
\email{Martin.Kunz@physics.unige.ch}
\affiliation{D\'epartement de
Physique Th\'eorique, Universit\'e de
Gen\`eve, 24 quai Ernest Ansermet, CH--1211 Gen\`eve 4, Switzerland}
\author{Bruce A. Bassett}
\email{bruce@saao.ac.za}
\affiliation{South African Astronomical Observatory, Observatory, Cape Town, South Africa \\ 
Department of Mathematics and Applied Mathematics, University of Cape Town, Rondebosch, 7700, Cape Town, South Africa}
\author{Ren\'ee A. Hlozek}
\email{reneeh@nassp.uct.ac.za}
\affiliation{National Astrophysics and Space Science Programme\\ 
University of Cape Town, Rondebosch, 7700, Cape Town, South Africa}

\begin{abstract}
Observed data are often contaminated by undiscovered interlopers, leading 
to biased parameter estimation. Here we present BEAMS (Bayesian Estimation Applied to Multiple Species) 
which significantly improves on the standard maximum likelihood 
approach in the case where the probability for each data point being `pure' 
is known. We discuss the application of BEAMS to future Type Ia supernovae (SNIa) 
surveys, such as LSST, which are projected to deliver over a million supernovae lightcurves 
without spectra. The multi-band lightcurves for each candidate will provide a 
probability of being Ia (pure) but the full sample will be 
significantly contaminated with other types of supernovae and transients. Given a sample 
of $N$ supernovae with mean probability, $\langle P \rangle$, 
of being Ia, BEAMS delivers 
parameter constraints equal to $N\langle P \rangle$ 
spectroscopically-confirmed SNIa. In 
addition BEAMS can be simultaneously used to tease apart different 
families of data and to recover properties of the underlying distributions 
of those families (e.g. the Type Ibc and II distributions). Hence BEAMS 
provides a unified classification and parameter estimation methodology 
which may be useful in a diverse range of problems such as photometric 
redshift estimation or, indeed, any parameter estimation problem where 
contamination is an issue.
\end{abstract}

\maketitle

\section{Introduction}

Typically parameter estimation is performed with the assumption that all the data
come from a single underlying probability distribution with a unique
dependence on the parameters of interest. In reality the dataset is
invariably contaminated by data from other probability distributions
which, left unaccounted for, will bias the resulting best-fit
parameters. This is a typical source of systematic error.

In this paper we present BEAMS (Bayesian Estimation Applied to
Multiple Species), a method that allows for optimal
parameter estimation in the face of such contamination when the
probability for being from each of the distributions is known. As a
by-product our method allows the properties of the contaminating
distribution to be be recovered.

For example, the next decade will see an explosion of supernova data
with particular emphasis on Type Ia supernovae (SNIa) as standard
candles. A few hundred supernovae were known by 2005, see \cite{riess98, perlmutter99,hamuy9601, hamuy9602, riess99, tonry03, riess04} and references therein. The current generation of SNe surveys will last to around 2008 and include SNLS \cite{snls, 0510447},
ESSENCE\cite{essence, 0411357}, SDSS-II\cite{0504455}, \footnote{http://sdssdp47.fnal.gov/sdsssn/sdsssn.html}, CSP \cite{0512039}, \footnote{http://csp1.lco.cl/$\sim$cspuser1/CSP.html}, KAIT \footnote{http://astro.berkley.edu/$\sim$bait/kait.html}, CfA \cite{hicken06, jha06}, C-T\cite{ct} and SN Factory\cite{snfac} and will yield of order $10^3$ good SNIa with spectra. 
Proposed next-generation supernova surveys include the Dark Energy Survey \cite{des}, 
Pan-STARRS \cite{panstarrs} and SKYMAPPER\cite{skymapper} and will deliver of order 
$6 \times 10^4$ SNIa by 2013, the majority of which will {\em not} have spectra. 
Beyond this the projected ALPACA telescope \cite{alpaca} would find an estimated $10^5$ 
SNIa over three years. The exponential data rush will culminate in the LSST \cite{lsst}, \footnote{http://www.lsst.org} which is expected to discover around $2 \times 10^5$ SNIa per year, yielding a catalog with over two million SNIa multi-colour light curves over a ten year period. The vast majority of these candidates will not have associated spectra. 

Fortunately recent surveys such as HST, SNLS and SDSS-II \cite{hst, riess06, photo_sn1}, \footnote{http://sdssdp47.fnal.gov/sdsssn/sdsssn.html} building on earlier work 
have convincingly shown that a probability of any object being a SNIa can be derived from 
multi-colour photometric observations of the candidate. This has become a very active 
area of research with significant recent advances pursuing a primarily Bayesian 
approach to the problem \cite{sdss, prob, bayes_1, red} 
and suggesting that the future high-quality, multi-epoch lightcurves will provide 
accurate (i.e. relatively unbiased) probabilities of being each possible type of supernova (or of not being a supernova at all). 

However, since a less than $100\%$ probability of being Ia is insufficient for the standard 
parameter estimation methodology, these probabilities - no matter how accurate they are - 
are useless and have been relegated to use in selecting targets for spectroscopic follow-up 
as it has always been considered imperative to obtain spectra of the candidates to find Ia's, 
reject interlopers and to obtain a redshift for the SNIa. 

As a result, even with the relatively small number of supernova candidates today it is impossible to obtain spectra for all good potential SNIa candidates. 
Instead only the best candidates are followed up. For LSST and similar telescopes less 
than $0.1\%$ of likely SNIa candidates will be followed up spectroscopically.
Unfortunately a spectrum for a high-z object is typically very costly to obtain, with 
the required integration time roughly scaling as $(1+z)^\alpha$ with $\alpha$ somewhere between 2 and 6, depending on the specific situation. In practise the situation is more complex since key identifying features such as the Si II absorption feature at a rest frame $6150 \AA$ are redshifted out of the optical at $z \sim 0.4$ requiring either infra-red observations or higher signal-noise spectra of the remaining part of the spectrum.


\begin{figure}[ht]
\centerline{\epsfig{figure=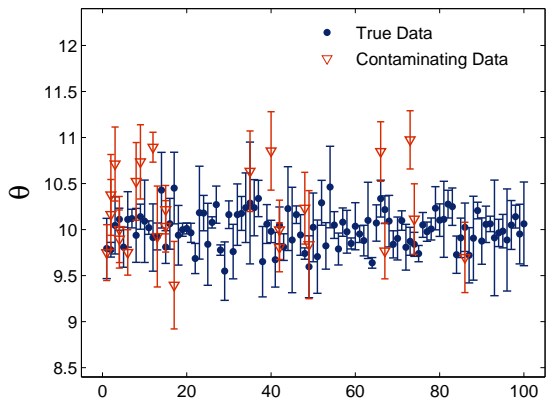,width=8cm}}
\caption{ \label{fig:schem_data} 
Schematic illustration of the problem: data drawn from the true distribution (e.g. 
Type Ia supernovae) are contaminated by similar looking data from a different 
distributions (e.g. Type Ibc or II supernovae) leading to biasing in the 
best-fit for parameters and in their errors.
}
\end{figure}

\begin{figure}[ht]
\centerline{\epsfig{figure=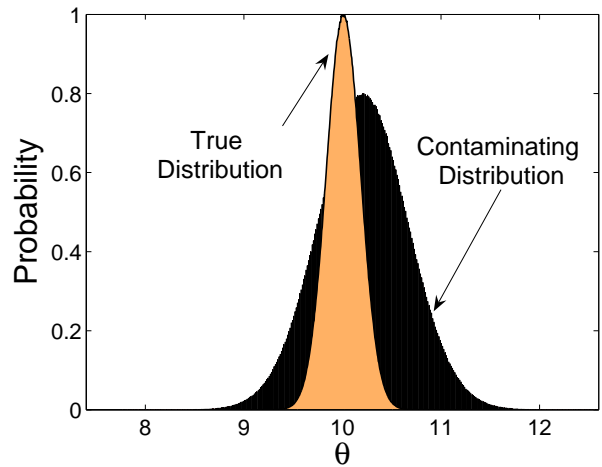,width=8cm}}
\caption{ \label{fig:schem_dist} 
Schematic illustration of the underlying distributions used in Figure (\ref{fig:schem_data})
for the true and contaminating data. In the case of supernovae, Type Ia have a much more narrow
intrinsic scatter in their intrinsic luminosity (narrow Gaussian) compared to other types of
supernovae (wide Gaussian) which also have different means. 
}
\end{figure}

Until now the choices available in dealing with such a flood of
candidates were limited. Either one could limit oneself to those
candidates with spectra, rejecting the vast majority of candidates, or
one could imagine using the full dataset - including the contaminating
data - to perform parameter estimation. However, undertaking this in a
naive way - such as simply accepting all candidates which have a
probability of being a SNIa greater than some threshold, $P_*$ - will
lead to significant biases and errors that will undermine the entire
dataset.

In contrast, we introduce in this paper a statistically rigorous method 
for using the candidates 
without spectroscopic confirmation for parameter estimation. 
BEAMS offers a fully Bayesian method for appropriately
weighting each point based on its probability of belonging to each
underlying probability distribution (in the above example, its
probability of being a SNIa, SNIbc, type II etc...). We will show that
this leads to a parameter estimation method without biases (as long as
the method for obtaining the probabilities is sound) and which improves
significantly the constraints on (cosmological) parameters.

We will be 
guided by resolving this specific problem, but the underlying principles and methods 
are more general and can be applied to many other cases. In order not to obscure
the general aspects, we will skip over some details, leaving them for future work
where actual supernova data is analysed. We will therefore assume here that we know
the redshift of the supernovae (or of its host galaxy), and that we already have
estimated the probabilities $P_j$ that the j-th supernova is a SN-Ia (eg. by
fitting the lightcurves with templates).

To give a simple example, imagine that we wish to estimate a parameter
$\theta$ (which in cosmology could for example represent the
luminosity distance to a given redshift) from a single data point,
$D$, which could have come from one of two underlying classes
(e.g. supernova Type Ia or Type II), indexed by $\tau=A,B$ (with their
own probability distributions $P(\theta|D, \tau)$, for the parameter
$\theta$). Again considering SNe, the link between luminosity and the
luminosity distance could be different for the different classes of
supernovae due to their intrinsic distribution properties, so
given the data $D$, what is the posterior likelihood for $\theta$
assuming that we also know the probability, $P_{\tau}$ that the data
point belongs to each class, $\tau$?

Clearly, $P_A = 1-P_B$ since we assume the point could come from
only one of two classes. Secondly, as $P_{\tau} \rightarrow 0,1$, the
posterior should reduce to one or other of the class distributions. 
Hence by continuity, the posterior we are seeking should have the form:
\be
P(\theta|D) = f(P_A) P(\theta|D, \tau=A) + g(P_B) P(\theta|D, \tau=B)
\ee
where the continuous functions $f$ and $g$ have the limits 
$f(0) = 0 = g(0)$ and  $f(1) = 1 = g(1)$.
Since all the posteriors are normalised we have that  $\int P(\theta|D) d\theta = 1
= \int P(\theta|D,\tau) d\theta$. We immediately find that $g(P_A) = 1
- f(1-P_A)$.
 The simplest -- and as we will show later, Bayesian --
choice for $f$ is simply the linear function: $f(P_A) = P_A$.  In this
case the full posterior simply becomes: 
\be 
P(\theta|D) = P_A P(\theta|D,\tau=A) + (1-P_A) P(\theta|D, \tau=B) 
\ee 
This can be easily understood: the final probability distribution
for $\theta$ is a weighted sum of the two underlying probability
distributions (one for each of the classes) depending on the
probabilities $P_A,P_B (= 1 - P_A)$ of belonging to each of the two
classes.

We will see that our general analysis bears this simple intuition out 
(see e.g. equation (\ref{eq:full_post})).

\section{Formalism}

\subsection{General case}

Let us derive in a rather general way the required formulae. Starting
from the posterior distribution of the parameters, $P(\pp|\dat)$ we
can work our way towards the known likelihood by repeated application
of the sum and product rules of probability theory. The crucial first 
step involves writing
explicitly the marginalisation over different data populations,
represented by a logical vector $\pop$. Each entry $\pop_i$ is either
$A$ if the supernova $i$ is of type Ia, and $B$ if it is not. With
each entry we associate a probability $P_i$ that $\pop_i=A$, so that
the probability for $\pop_i=B$ is $1-P_i$. For now we assume that
these probabilities are known. We can then write
\be
P(\pp|\dat) = \sum_\pop P(\pp,\pop|\dat)
\ee
where the sum runs over all possible values of $\pop$. Using Bayes
theorem we get
\be
P(\pp,\pop|\dat) = P(\dat|\pp,\pop) \frac{P(\pp,\pop)}{P(\dat)} .
\label{eq:bayes}
\ee
The ``evidence'' factor $P(\dat)$ is independent of both the
parameters and $\pop$ and is an overall normalisation that can
be dropped for parameter estimation. We will further assume
here that $P(\pp,\pop) \approx P(\pp) P(\pop)$. This simplification
assumes that the actual parameters describing our universe are
not significantly correlated with the probability of a given
supernova to be of type Ia or of some other type. Although it is
possible that there is some influence, we can safely neglect it
given current data as our parameters are describing the large-scale
evolution of the universe, while the type of supernova should mainly
depend on local gastrophysics. In this case $P(\pp)$ is the usual
prior parameter probability, while $P(\pop)$ separates into independent
factors,
\be
P(\pop) = \prod_{\pop_i=A} P_i \prod_{\pop_j=B} (1-P_j) ,
\ee
Here the product over ``$\pop_j=A$'' should be interpreted 
as a product over those $j$ for which $\pop_j=A$. In other words, given
a population vector $\tau$ with entries ``$A$'' for SN-Ia and ``$B$'' for
other types, the total probability $P(\pop)$ is the product over all entries, 
with a factor $P_j$ if the j-th entry is ``$A$'' and $1-P_j$ otherwise (if the
j-th entry is ``$B$''). Notice that we discuss here
only one given vector $\pop$, the uncertainty is taken care of by
the outer sum over all possible such vectors. The full expression
is therefore
\be
P(\pp|\dat) \propto P(\pp) \sum_\pop P(\dat|\pp,\pop)  
\prod_{\pop_i=A} P_i \prod_{\pop_j=B} (1-P_j). \label{eq:posterior}
\ee
The factor $P(\dat|\pp,\pop)$ here is just the likelihood.
In general we have to evaluate this expression, which is composed
of $2^N$ terms for $N$ supernovae. The exponential scaling with
the number of data points means that we can in general not evaluate
the full posterior --  but it should be sufficient to fix $\tau_i=A$ for
data points with $P_i \approx 1$ and $\tau_j=B$ for $P_j \approx 0$,
and to sum over the intermediate cases. This should give a sufficiently
good approximation of the the actual posterior.

\subsection{Uncorrelated data}
In the case of uncorrelated kinds of data or measurements, such as is approximately true for
supernovae \footnote{Note that this is an idealisation for supernovae since at low redshifts supernovae
are correlated due to large-scale bulk velocity fields \cite{velocity,0603240, 0512159}. 
Further, if the supernovae hosts have redshifts estimated from photometry instead of spectroscopy then correlations between SNe will be induced when the $4000\AA$ break lies in the same filter and will be exacerbated by host extinction issues, see e.g. \cite{alex1,dragan}. Since we wish to present the general formalism here we assume that we know the redshifts of the SNe perfectly. In general the estimation of the redshift must be included in the parameters to be estimated from the data. In addition to these challenges, the template correction is usually
computed using the supernova sample itself and may introduce some
correlations. But in general, if the property of supernovae which makes
them standard candles depends {\em on the sample} then we should be
worried. So here we assume that the template corrections and errors
were derived previously with the spectroscopic supernovae. Indeed, strictly
speaking, we should use a {\em different} sample for that purpose, or else
estimate those parameters as well as the global dispersion {\em simultaneously}
with the cosmological parameters. In the latter case it is important to keep 
the $1/\sigma$ normalisation of the likelihood and to use additionally a
``Jeffreys prior'' $\propto 1/\sigma$ to avoid a bias towards larger dispersions. In the case where correlations are important then one must compute the full probability which is computationally intense, though systematic perturbation theory may be useful for small correlations.}, we 
can apply the huge computational simplification pointed out in
\cite{press}. In this case, the likelihood decomposes into a product
of independent probabilities,
\be
P(\dat|\pp,\pop) = \prod_{\pop_i=A} P(\dat_i|\pp,\pop_i=A)
\prod_{\pop_j=B} P(\dat_j|\pp,\pop_j=B) .
\ee
The posterior is now a sum over all possible products indexed by the components $\tau_i$. 
We can simplify it, and bring it into a form that
lends itself more easily to the extensions considered in a later
section, by realising that all binomial
combinations can be generated by a product of sums of
two terms,
\be
\sum_{\pop} \prod_{\tau_i=A} A_i \prod_{\tau_j=B} B_j
= \prod_k (A_k+B_k) .
\ee
In this schematic expression, the $A_i$ correspond to the product
of likelihood and prior for a $\tau_i=A$ entry, and the $B_j$
to the same product for a $\tau_j=B$ entry. So instead of a sum
over $2^N$ terms, we now only deal with $N$ products.

How do the $A_k$ and $B_k$ look for our supernova application?
Let us assume that we are dealing with two populations, a population
$A$ of SNe Ia and population $B$ of non-Ias. For the $k$-th supernova
$A_k$ is then the product of the probability $P_k$ of being type Ia
with the likelihood $P(\dat_k|\pp,\pop_k=A)$. But since this likelihood
is conditional on the supernova being indeed of type Ia, it is just
the normal type-Ia likelihood which we will call $\LL_{A,k}$. $B_k$ on
the other hand is the probability $1-P_k$ of not being type Ia times
the likelihood of the supernovae that are not Ia, which we will call
$\LL_{B,k}$.

$\LL_{A,i}$ is therefore the probability that the $i$-th data point
has the measured magnitude if it is type Ia. It is just the
usual likelihood, typically taken as a $\chi^2$ in the magnitudes.
With the $i$-th supernova data given as distance modulus $\mu_i$
and total combined error $\sigma_i$ (the intrinsic and measurement errors computed in quadrature) it is simply
\be
P(\dat_i|\pp,\pop_i=1) = {\cal L}_{A,i}(\pp) = \frac{1}{\sqrt{2\pi}\sigma_i}
e^{- \chi_i^2 / 2},
\ee
with $\chi_i^2 = (\mu_i - m(\pp))^2/\sigma_i^2$ where
$m(\pp)$ is the theoretical distance modulus (at redshift $z_i$).
We emphasise that here the normalisation of the likelihood is
important -- unlike in standard maximum likelihood parameter estimation -- as we will be dealing with different distributions and their relative weight depends on the overall normalisation. In the case of SNe we can of course go a level deeper, 
since the $\mu_i$ are estimated from a number of light-curve points in multiple filters. We could start directly with those points as our fundamental data. Here we ignore this complication while noting that in an actual application this would be the optimal approach \footnote{We thank Alex Kim for pointing this out to us.}

The likelihood $\LL_{B,i}$ of a non-Ia supernovae is harder. In an ideal world
we would have some idea of the distribution of those supernovae,
so that we can construct it from there (see e.g.~\cite{richardson}). 
If we do not know anything,
we need to be careful to minimise the amount of information that
we input. It is tempting to use an infinitely wide flat distribution,
but such a distribution is not normalisable. Instead we can assume
that the non-Ia points are offset with respect to the ``good'' data
and have some dispersion. The natural distribution given the first
two moments (the maximum entropy choice) is the normal (Gaussian) 
distribution. The potentially most elegant approach
is to use the data itself to estimate the width and location of
this Gaussian. This is simply done by allowing for a free 
shift $b$ and width $\Sigma$ and marginalising over them. 
Optimally we should choose both parameters independently for each redshift
bin, in the case where we have many supernovae per bin.
Otherwise it may be best consider $b$ as a relative shift with
respect to the theoretical value, modelling some kind of bias. 

We would like to emphasise that our choice of the normal distribution
for the non-Ia points is the conservative choice if we want to add a
minimal number of new parameters. It does not mean that we assume it
to be the correct distribution. In tests with a uniform and a $\chi^2$
type distribution for the non-Ia population, assuming a normal distribution
sufficed to reliably remove any bias from the estimation process relying on 
the Ia data points. If we have a very large number of non-Ia points
we could go beyond the normal approximation and try to estimate the
distribution function directly, e.g. as a histogram. On the other hand,
the more parameters we add, the harder it is to analyse the posterior.
Also, if we {\em knew} the true distribution of the contaminants
then we should of course use this information.
Going back to the full likelihood, we now write \footnote{Again we stress that for the sake of clarity and generality we assume that we know the redshifts of the SN perfectly. If the redshift must also be estimated from the data then the formalism below must be extended in the obvious way.}
\bea
P(\dat|\pp,\pop) 
&=& \sum_{b,\Sigma} P(\dat,b,\Sigma|\pp,\pop) \\
&=& \sum_{b,\Sigma} P(\dat|b,\Sigma,\pp,\pop) P(b,\Sigma) .
\eea
The last term is the prior on the non-Ia distribution. In the
absence of any information, the conventional (least informative)
choice is to consider
the two variables as independent, with a constant prior on $b$ and
a $1/\Sigma$ prior on the standard deviation. In reality, the sum written
here is an integration over the two parameters, and the choice
of prior is degenerate with the choice of integration measure.
As there are no ambiguities, we will keep using summation
symbols throughout, even though they correspond to integrals
for continuous parameters. 

The type-Ia supernovae are independent of the new parameters. 
They are only relevant for the non-Ia
likelihood, which is now for supernova $j$
\bea
P(\dat_j|\pp,b,\Sigma,\pop_j=B) &=& {\cal L}_{B,j}(\pp,b,\Sigma) \nonumber \\
&=&\frac{1}{\sqrt{2\pi}\Sigma}
e^{-\frac{(\mu_j-m(\pp)-b)^2}{2\Sigma^2}} 
\eea
(in an actual application to supernova data we would take
$\Sigma$ to be the intrinsic dispersion of the non-Ia population,
and add to it the measurement uncertainty in quadrature).
The posterior, Eq.~(\ref{eq:posterior}), is then
\bea
P(\pp|\dat) \propto \sum_{b,\Sigma} P(\pp) P(b) P(\Sigma)\times \nonumber\\
\prod_{j=1}^N \left\{ \LL_{A,j}(\pp) P_j + \LL_{B,j}(\pp,b,\Sigma)(1-P_j)  \right\}
\label{eq:full_post}
\eea
An easy way to implement the
sum over $b$ and $\Sigma$ is to include them as normal variables
in a Markov-chain monte carlo method, and to marginalise over
them at the end. Additionally, their posterior distribution
contains information about the distribution of the non-Ia
supernovae that can be interesting in their own right.



\section{A Test-Implementation\label{sec:testmodel}}

In general $\theta$ could of course be a vector of cosmological
parameters, but in this section we consider the simple case of the
estimation of a constant, corresponding for example to the luminosity
distance in a single bin for the SN case. Continuing with the SN
example for simplicity, the data $D_i$ then corresponds to some $m_i$,
an apparent magnitude for each SN in a bin. We again assume that there
are two populations, type $A$ (corresponding to SNIa) and type $B$
(everything else).

We fix a distribution for the type $A$
probabilities $P_i$, for simplicity we take $f(P_i)\propto P_i$, i.e.
a distribution that is linearly increasing so that we are dealing
predominantly with objects of type $A$. We then draw a $P_i$ from this
distribution, and choose an actual type with that probability.
Finally, we add a ``spectroscopic'' sample for which $P_i=1$, i.e. these
are guaranteed to be of type $A$.

We take the type $A$ population to have a known Gaussian distribution 
with mean $\mu_A=0$ and variance $\sigma_A=0.1$. The unknown distribution
of type $B$ is taken to be another Gaussian, with mean $\mu_B=2$ and
variance $\sigma_B=2$. To all data points, $A$ and $B$, we assign the error bar of
type $A$, i.e. $\sigma_i=\sigma_A$ (but we fit for the error bar of the population $B$). 
We assume that this error has been derived
e.g. from the dispersion of the spectroscopic sample and that we do not
know the distribution of the sample $B$ \footnote{Although this specific example considers a 
Gaussian distribution for the $m_i$ of the ``non-Ia'' population $B$, which corresponds
its likelihood, we have also tested the algorithm for other distributions.}.

The parameters that are being fitted from the data are then $\mu_A$,
$\mu_B$ and $\sigma_B$, with $\sigma_A$ fixed from the spectroscopic
sample, and $P_i$ fixed for each point from an assumed previous step
in the analysis (e.g. $P_i$ obtained from goodness of fit to template
lightcurves). As a side remark, although $\sigma_A$ is here assumed to be known
from the dispersion of the spectroscopic sample, it can also be fitted for 
jointly with the other
parameters, which was done in tests of the method \footnote{In this case
its prior needs to be $\propto1/\sigma_A$ to avoid biases.}; the assumption of
fixed known $P_i$ will be relaxed in later sections.
To connect this highly simplified example with cosmology, we shall pretend that we consider
here only one redshift bin, and that the same analysis is repeated for
each bin. The value of $\mu_A$ could then be the distance modulus $\mu$
in one bin, and an unbiased estimate in all bins 
 would then constrain cosmological parameters like 
$\Omega_m, \Omega_{\Lambda}$ etc. The smaller the errors on $\mu_A$,
the better the constraints. The data from population
$B$ on the other hand give us no information on the distance modulus, 
hence we must reduce
contamination from population $B$. The posterior that results
(explicitly indicating that we estimate $\mu_A$) is then 
\bea
P(\mu_A|\dat,\sigma_A) \propto \sum_{\mu_B,\sigma_B} \frac{1}{\sigma_B} \times
\nonumber\\ \prod_{j=1}^N [P_j {\cal L}_{A, j}(\mu_A,\sigma_A) + (1-P_j) 
{\cal L}_{B, j}(\mu_B,\sigma_B)], 
\label{eq:test}
\eea
where the population $B$ mean $\mu_B$ and the variance $\sigma_B$ have taken over the
role of the shift $b$ and variance $\Sigma$ of the last section.

As the population $B$ is strongly
biased with respect to $A$, the algorithm needs to detect the type correctly
to avoid wrong results. Table \ref{tab:test1} shows results
from an example run with the above parameters, $10$ spectroscopic and $1000$
photometric data points, where the spectroscopic points are data generated in a Monte Carlo fashion from normally distributed population $A$ and the photometric data consist of points from both population $A$ and population $B$ with associated probabilities $P_i \le 1$. 
In this table and all following tables we add a ``Bias'' column that shows
the deviation of the recovered parameters from the input values in units
of standard deviations.

\begin{table}
\centering
\begin{tabular}{ccc} \hline
Parameter &  Value & Bias [$\sigma$]\\ \hline
$\mu_A$ & $-0.003\pm0.004$ & $0.8$ \\
$\mu_B$ & $2.00\pm0.11$ & $0.0$ \\
$\sigma_B$ & $1.90\pm0.07$ & $1.4$\\
\hline
\end{tabular}
\caption{Example results for the basic algorithm
applied to a sample of $10$ ``spectroscopic'' and
$1000$ ``photometric supernovae'' in a bin. The bias
column shows the deviation from the true value, in units
of the standard deviation. A deviation of about $1 \sigma$
is expected, while about one in twenty realisations is
more than $2 \sigma$ away for random data with normal
distribution. BEAMS also allows to recover the
parameters characterising the contaminating distribution,
$\mu_B$ and $\sigma_B$.}
\label{tab:test1}
\end{table}

\begin{figure}[ht]
\centerline{\epsfig{figure=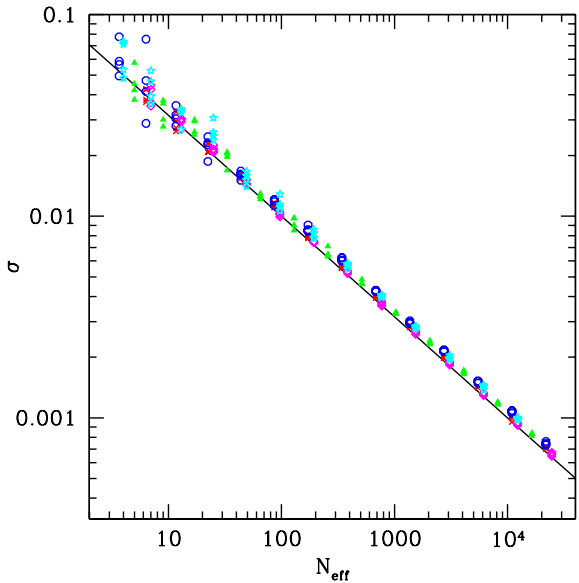,width=8cm}}
\caption{ \label{fig:error} Scaling of the errors: The black line
shows the expected (optimal) error, which is inversely proportional
to the effective number of SNIa given by $\sqrt{N_\eff}$ in eq. (\ref{eq:neff}). The different colours and shapes correspond to different distributions of the probabilities $P_i$ (i.e. how many data points have $P_i=0.9$, how many have $P_i=0.8$, etc.). The points show the actually measured error for BEAMS given these distributions of the probabilities $P_i$ of the data. BEAMS is able to use nearly
all of the information available.
}
\end{figure}

For the spectroscopic sample the errors just scale like $\sigma_A/\sqrt{N}$.
Each of the other supernova contributes to the ``good'' measurement with
probability $P_j$, i.e. each data point has a weight $P_j$, or an effective
error bar $\sigma_A/\sqrt{P_j}$ on average. Defining the average weight
\be
w\equiv \frac{1}{N} \sum_{j=1}^N P_j \rightarrow \int d\!P P f(P)
=\langle P \rangle
\ee
where $f(P)$ is the normalised probability
distribution function of the $P_j$, we find that the error on $\mu$ 
scales as
\be
\sigma_\mu = \frac{\sigma_A}{\sqrt{N_s + w N_{ph}}}
\ee
for $N_s$ spectroscopic measurements ($P_j=1$) and $N_{ph}$ uncertain (photometric only) measurements with average weight $w$. As can be seen in Fig.~\ref {fig:error},
the errors on $\mu$ recovered by the Bayesian formalism
do indeed follow this formula, although they can be slightly worse if
the two populations are more difficult to separate than in this example.

In our example where $f(P_j)\propto P_j$ the weight is $w=2/3$, so that 
three photometric supernovae
equal two spectroscopic ones. The expected error in $\mu_A$ for the example of
table \ref{tab:test1} is therefore $0.1/\sqrt{10+2/3\times 1000}\approx0.004$,
in agreement with the numerical result. If we had used only the $10$ spectroscopic
data points, the error would have been $0.032$ so that the use of all available
information improves the result by a factor eight.
In the case where $f(P_i) \propto (1-P_i)$,
i.e. we are dealing predominantly with type $B$ data, we have a weight of
$1/3$. If it is easier to measure three photometric supernovae compared to
one spectroscopic one, it will still be worth the effort in this case.
We should point out here that these are the optimal errors achievable with
the data. In Fig.~\ref{fig:error} we show the actual recovered error from
random implementations with different $w$ and effective number of SNIa given by:
\be
N_\eff\equiv N_s + w N_{ph}\,.
\label{eq:neff}
\ee
We see that the Bayesian algorithm achieves nearly optimal errors
(black line). 

We now compare the Bayesian method to some other possible methods:
\begin{itemize}
\item Use only spectroscopic SNIa.
\item Use only SNIa with probabilities above a certain
limiting threshold, $P$. A limit of $0\%$ uses all data points, and
a limit of $100\%$ only the spectroscopically confirmed points.
\item Weight the $\chi_i^2$ value for the $i$-th point by a function of $P_i$. This effectively 
corresponds to increasing the error for data points with lower probability.
For the test, we use the weighting $\sigma_j \rightarrow \sigma_j/P_j^{N/2}$. For $N=0$ 
this reverts to the limiting case where we just use all of the data
in the usual way. For $N > 0$ points are progressively more and more heavily penalised for having low probabilities.
\end{itemize}
These ad-hoc prescriptions are not necessarily the only possibilities, but these were the methods we came up with
for testing BEAMS against. We now discuss their application to the same
test-data described above to see how they perform against BEAMS.

\begin{figure}[ht]
\centerline{\epsfig{figure=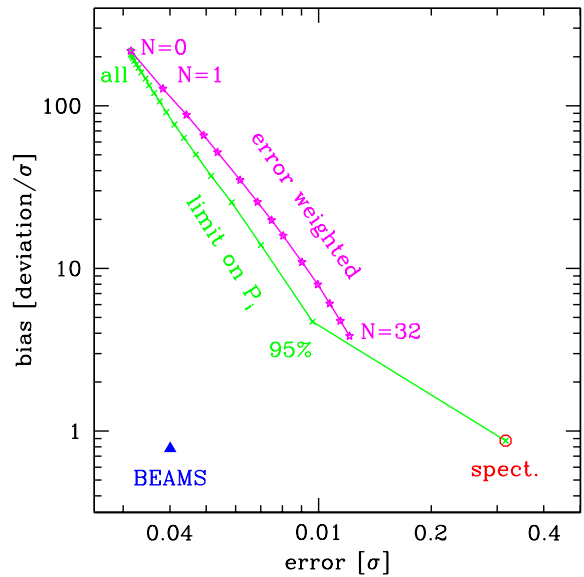,width=8cm}}
\caption{ \label{fig:comparison} A comparison between different
methods (see text). Of these methods, only the BEAMS method
and the use of the spectroscopic points alone are unbiased. As it
can use the uncertain data, the BEAMS method improves the
error bars in this example by the (expected) factor of $8$.
(For the error-weighted method not all values of $N$ were
plotted at the high-N end.)
}
\end{figure}

Figure \ref{fig:comparison} shows very clearly that although the
ad-hoc prescriptions for dealing with the type-uncertainty can lead to
very precise measurements, they cannot do so without being very
biased. Both the Bayesian and the pure-spectroscopic approach recover
the correct value (bias less than one $\sigma$), but the latter does
so at the expense of throwing away most of the information in the
sample.

\begin{figure}[ht]
\centerline{\epsfig{figure=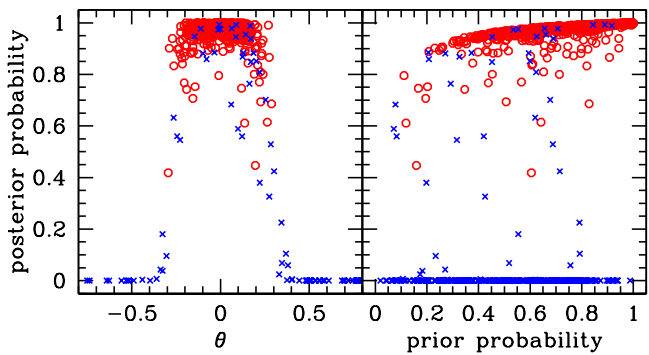,width=9cm}}
\caption{\label{fig:postprob} BEAMS as a classification algorithm:
we plot the posterior probability of the data points to be of type
$A$ in our toy example. This depends on their value $m_j$ (left panel) 
and their prior probability (right panel). The better a point agrees
with the recovered distribution of type $A$, the higher its posterior
probability to belong to it. See text for more information. For this test case we know the true nature of the points,
and plot population $A$ as red circles and population $B$ as blue crosses (see Table \ref{tab:test1} for distribution characteristics). . 
}
\end{figure}

We can also use BEAMS to get a posterior estimate of the population
type, based on the prior value (e.g. from multicolour light curves)
and the distribution. To do this for data point $j$ we marginalise
over all entries $\pop$ {\em except} $\tau_j$, and additionally over
all estimated parameters. In practice this means that the $j-th$ entry
in Eq.~(\ref{eq:test}) is fixed to $\LL_{A,j}$, and that we also
integrate over $\mu_A$ in addition to $\mu_B$ and
$\sigma_B$. Effectively, we compute the model probability if the
$j$-th point is assumed to be of type $A$ and compare it to the model
probability without this constraint.  The relative probability of the
two cases then tells us the posterior probability for the model vector
$\pop$ having the $j$-th entry equal to $A$, corresponding to the
posterior probability of the $j$-th supernova to be of type
$A$. Fig.~\ref{fig:postprob} shows an example case (using a Gaussian
approximation to evaluate the integral over all values of the sample
mean $\mu_A$). We see how the posterior probability to belong to
population $A$ depends both on how well the location of a point agrees
with the distribution of $A$ (left) and on how high its prior
probability was (right). In other words, we can reconstruct which
points came from which distribution from the agreement between their
values of $\mu$ and $\mu_A$ and their prior probabilities (which is
indeed all the information at our disposal in this scenario).

For the toy example the two distributions are quite different, and BEAMS classifies all points
within about $3\sigma$ of $m=0$ to be of population $A$. The prior
probability is here strongly overwhelmed by the data and the resulting
posterior probabilities lie close to $0$ and $1$ for most data points.

In the following section we extend this basic model in two main directions.
Firstly, as reality starts to deviate from the model, there is a danger
of introducing a bias. We discuss a few simple cases and try to find
ways of hardening the analysis against the most common problems.
Secondly, we extend the model to more than two families, and we also
discuss the possibility of using the information on the other populations
in the analysis itself.

\section{Extensions}

\subsection{Uncertain probabilities}

While the likelihoods used in the estimation of $\mu_A$ (which will yield $\theta$) are the same for the earlier example, in this section our treatment and use of the probabilities $P_i$ differs as we begin to include possible error in the $P_i$'s.

Often one may not know the population probability $P_j$ precisely, but
has instead a probability distribution. For example $P_j$ may be roughly
known, but has an error associated with it (in the SN case this could be due to some systematics in the lightcurve fitting process). In this case we have to
marginalise over all those probability distributions. For $N$ supernovae
this then requires an $N$-dimensional integration. It is straightforward
to include this in a MCMC approach by allowing all $P_j$ to be free
variables, but with $N$ of the order of several thousand it may be
difficult to get a precise result. On the other hand this may still be
better than just sampling $P_j$ at a single point if it is not known
exactly.

However, if the measurements are independent, then
each integral affects only one of the terms in the product over
all data points in Eq.~(\ref{eq:full_post}). Instead of one $N$ dimensional
integration we are dealing with $N$ one-dimensional integrations which are
much easier to compute. In general we have to integrate each term
over the probability $p_j$ with a given distribution $\pi(p_j)$. The 
case of a known probability corresponds to $\pi(p_j) = \delta(p_j-P_j)$.
The next simplest example is the case
of a totally unknown probability $P_j$, for which $\pi(p_j) = 1$. In 
this case the integral to be solved in each term is
\be
\int dp_j \left({\cal L}_{A,j} p_j + {\cal L}_{B,j} (1-p_j)  \right) = 
\frac{1}{2} ({\cal L}_{A,j}+{\cal L}_{B,j}) ,
\ee
where ${\cal L}_{A,j}$ and ${\cal L}_{B,j}$ are the likelihood values 
of the $j$-th data point assuming population $A$ or $B$ respectively.
The effective probability here turns out to be $P_j=1/2$. The reason
is that we estimate this probability independently for each supernova,
and do not have enough information to estimate it from the data. In
the following subsection we replace this approach instead with a global
uncertain probability added to the known distributions. This global
probability can then be estimated from the data.

For now, assume we have an approximate knowledge of the type-probabilities,
say, an independent uncertainty on each $P_j$, $\delta_j$, so that
\be
\pi(p_j) \propto e^{-\frac{(p_j-P_j)^2}{2 \delta_j^2}} ,
\ee
where the proportionality constant is chosen so that the integral
over $\pi(p_j)$ from zero to one is $1$. If the random error on $P_j$
is small enough that the distribution function is well contained within
the domain of integration, i.e. $P_j+\delta_j\ll1$ and $P_j-\delta_j\gg0$,
then we recover just $p_j=P_j$. In this case the Gaussian distribution
function acts effectively as a delta-function. For large uncertainties,
or for probabilities close to the boundaries, corrections will become
important and can bias the result. For the specific case of random
errors, the correction term is of the form 
$\LL_{A,j}-\LL_{B,j}$. If we suspect large
{\em random} errors it may be worth adding this term with a global
pre-factor of its own to the full posterior. On the other hand, in real
applications we expect that the probabilities close to $P_j=1$ are
quite well known, so that the boundary error is hopefully not too important.

A fixed, common shift is much more worrying and can bias the results
significantly. This can be seen in table \ref{tab:test2} where we added
a systematic shift to the probabilities (enforcing $0\leq P_j\leq1$).
This is an especially important point for photometric supernova analyses,
where dust reddening can bias the classification algorithm. If we do not
take into account this possibility, then the analysis
algorithm fails because it starts to classify the supernovae wrongly,
but hopefully such a large bias is unrealistic.
\begin{table}
\centering
\begin{tabular}{ccc} \hline
shift of $P_j$ &  $\mu_A$  & Bias [$\sigma$]\\ \hline
$+0.1$ &  $0.021\pm0.004$  & $~5.5$  \\
$+0.2$ &  $0.128\pm0.004$  & $31.8$  \\
$+0.4$ &  $0.408\pm0.004$  & $96.8$  \\
$-0.4$ &  $0.003\pm0.005$  & $~0.6$ \\
\hline
\end{tabular}
\caption{Results with a systematic shift (i.e. bias) in the probabilities
$P_j$. Positive shifts lead to a systematic bias in the results,
while negative shifts lead to sub-optimal errors. However, the
negative shifts will bias instead the inferred properties of
population B.}
\label{tab:test2}
\end{table}

At any rate, a bias is readily dealt with
by including a free (global) shift $s$ into the probability factors of Eq.~(\ref{eq:full_post})
and by marginalising over it, resulting in 
\bea
P(\mu_A,\mu_B,\sigma_B|\dat) \propto \nonumber \\
\sum_s \frac{1}{\sigma_B} 
 \prod_{j=1}^N \left[ {\cal L}_{A,j}(P_j+s) + {\cal L}_{B,j}(1-P_j-s)\right].
\label{eq:full_post_shift}
\eea
It may be a good idea to include such a shift and to check its posterior distribution. Given enough
data it does not significantly impact the errors, and it adds stability
also in the case of large random uncertainties in the $P_j$. We found
that an additive bias with a constant prior was able to correct all
biasing models that we looked at, as is shown in table \ref{tab:test3}.  However, the
presence of a significant shift would indicate a failure of the
experimental setup and should be taken as a warning sign.

A free individual shift is degenerate
with the case of random uncertainties above, as it cannot be 
estimated from the data, and is not very useful in this context.

\begin{table}
\centering
\begin{tabular}{cccc} \hline
shift of $P_j$ &  $\mu_A$  & Bias [$\sigma$]& recovered shift \\ \hline
$+0.0$ &  $-0.003\pm0.004$ & $-0.8$ & $\hpm 0.002\pm0.011$  \\
$+0.1$ &  $-0.004\pm0.004$ & $-1.0$ & $\hpm 0.073\pm0.012$  \\
$+0.2$ &  $-0.000\pm0.004$ & $-0.1$ & $\hpm 0.158\pm0.015$  \\
$+0.4$ &  $-0.002\pm0.004$ & $-0.6$ & $\hpm 0.286\pm0.016$  \\
$-0.4$ &  $\hpm 0.004\pm0.004$ & $\hpm 1.0$ & $-0.396\pm0.013$ \\
\hline
\end{tabular}
\caption{Same as table \ref{tab:test2}, but the model allows for
a bias (shift) in the $P_j$. As most supernovae are population A, with
$f(P_j) \propto P_j$, the recovered shift grows slower than the input
shift. However, it still removes any bias in the estimation of $\mu_A$.}
\label{tab:test3}
\end{table}

\subsection{Global uncertainty}

Given how critical the accuracy of the type-probability $P_j$ is
in order to get correct results, it may be preferable, 
as an additional test, to discard this information completely.
This helps to protect against wrongly classified outliers and
the unexpected breakdown or biasing of the classification algorithm.

Even if the probability for a supernova to be either of type Ia or of
another type is basically unknown, corresponding to a large error
on all the $P_i$, not all is lost. We can instead include a global
probability $p$ that supernovae belong to either of the groups, and
then marginalise over it. In this way, the data will pick out the
most likely value for $p$ {\em and} which observations belong to
which class. In terms of the posterior (\ref{eq:full_post}) this
amounts just to replacing all $P_j$ with $p$ and to marginalise
over it,
\bea
P(\mu_A,\mu_B,\sigma_B|\dat) \propto \nonumber\\ 
\sum_{p} P(p) \frac{1}{\sigma_B} \prod_{j=1}^N \left\{ \LL_{A,j} p 
+ \LL_{B,j}(1-p)  \right\} .
\eea
The prior on $p$, $P(p)$, contains any knowledge that we have
on the probability that any given supernova in our survey is
of type Ia. If we do not know anything then a constant prior
works well.
As this is a global probability (i.e. all supernovae have the same
$p$), we cannot in this
form include any ``per supernova'' knowledge on $p$, gained for
example from spectra or light curves. For this we need to revert
to the individual probabilities discussed previously. However,
it is a good idea to include the spectroscopic (known to be good)
points with an explicit $p=1$ as they then define which population
is the ``good'' population and generally make the algorithm more
stable.

\begin{table}
\centering
\begin{tabular}{cccc} \hline
shift of $P_j$ &  $\mu_A$  & Bias [$\sigma$]& global probability \\ \hline
$+0.0$ &  $-0.003\pm0.004$ & $-0.8$ & $0.66\pm0.02$  \\
$+0.1$ &  $-0.004\pm0.004$ & $-0.9$ & $0.68\pm0.02$  \\
$+0.2$ &  $\hpm 0.000\pm0.004$ & $\hpm 0.0$  & $0.66\pm0.02$  \\
$+0.4$ &  $-0.003\pm0.004$ & $-0.7$ & $0.64\pm0.02$  \\
$-0.4$ &  $\hpm 0.004\pm0.004$ & $\hpm 0.9$ & $0.65\pm0.02$ \\
\hline
\end{tabular}
\caption{Same as table~\ref{tab:test2}, but the model uses an
estimated global probability $p$ for all supernovae and does not 
use the $P_j$ (so in reality all runs above are the same). The 
expected global probability is $p=N_\eff/N\approx0.66$.}
\label{tab:test4}
\end{table}

In our numerical tests with the toy model described in section \ref{sec:testmodel}
this approach works very well, see table \ref{tab:test4}. However if the two distributions are 
difficult to separate, with similar average and dispersion, then the
algorithm can no longer distinguish between them and concludes that the
data is compatible with having been drawn from a single distribution
with averaged properties.  This does normally not lead to a
high bias, since otherwise the data would have been sufficient
to tease the populations apart. Nevertheless, it seems
preferable to use the relative probabilities for the supernova types
when the information is available and reliable.

\subsection{Several populations}

For an experiment like the SDSS supernova survey, a more conservative
approach may be to add an additional population with a very wide error bar
that is designed to catch objects that have been wrongly classified as
supernovae, or those which got a very high Ia probability by mistake.

Of course there is no reason to limit ourselves to two or three populations,
given enough data. If we end up with several thousand supernovae per bin
we can try to use the data itself to understand the different
sub-classes into which the supernovae can be divided.

The expression (\ref{eq:full_post}) can be straightforwardly 
generalised to $M$ classes $A_i$
of objects (for example supernova types) with their own means
$\mu_i$ and and errors $\sigma_i$ as well as the probability for
data point $j$ to be in class $A_i$ of $P_j^i$,
\be
P(\mu_i,\sigma_i|\dat) \propto 
\frac{1}{\prod_{i=1}^M \sigma_i}
\prod_{j=1}^N \left\{ \sum_{i=1}^M \LL_{i,j}(\mu_i,\sigma_i) 
P_j^i\right \} . \label{eq:multipop}
\ee
For each data point $j$ the probabilities have to satisfy
$\sum_i P_j^i = 1$. Of course there has to be at least one
class for which the model is known, i.e. for which we know
the connection between $\mu_i$ and the (cosmological)
parameter vector $\theta$
(the ``Ia'' class in the supernova example), or
else it would not be possible to use this posterior for
estimating the model parameters $\theta$ and we end up with
a classification algorithm instead of constraining cosmology.

It is possible that we even do not know how many different populations to
expect. In this case we can just keep adding more populations to the
analysis. We should then also compute the evidence factor as a function
of the number $M$ of populations, $P(\dat|M)$, by marginalising 
the posterior of Eq.~(\ref{eq:bayes}) over
the parameters,
\be
P(\dat|M) = \sum_{\pp,\pop} P(\dat|\pp,\pop) P(\pp,\pop) .
\ee
This is just the integral over all $\mu_i$ and $\sigma_i$ of the ``posterior'' that we
have used so far, Eq.~(\ref{eq:multipop}), since we did not normalise it. 
Once we have computed this factor, then we can compare the relative
probabilities of the number of different populations by comparing their
evidence factor, since by Bayes theorem (again),
\be
P(M|\dat) = P(\dat|M) \frac{P(M)}{P(\dat)} .
\ee
The relative probability of models with $m_1$ and $m_2$ populations
is then
\be
\frac{P(\dat|m_1)}{P(\dat|m_s)} \frac{P(m_1)}{P(m_2)}
\ee
and usually (in absence of additional information) the priors are taken
to be $P(m_1)=P(m_2)$ so that the evidence ratio gives directly the
relative probability.

\subsection{Combined Formula}

What is the best way to combine the above approaches for future supernova
surveys? There is probably no ``best way''. For the specific example of
the SDSS supernova survey the probabilities for the different SN populations
are derived from $\chi^2$ fits to lightcurve templates \cite{photo_sn1}. We expect three
populations, Ia, Ibc and II, and objects that are not supernovae at all.
We expect that last class to be very inhomogeneous, but we would like 
to keep the supernovae. From the spectroscopically confirmed supernovae
we can learn what the typical goodness-of-fit of the templates is expected
to be and so calibrate them. Supernovae where the $\chi^2$ of all fits
is, say, $10$ higher than for the typical spectroscopic cases are discarded.
For the reminder we set $\pi_i = \exp(-(\chi_0^2-\chi^2)_i/2)$ where
$\chi_0^2$ is the typical value for each population. If $\sum_i \pi_i>1$
then we set the probabilities to be $P_j = \pi_j/\sum_i \pi_i$, otherwise
$P_j = \pi_j$. We also write again the more general $\theta$ for the 
parameters of interest. $\theta$ can represent for example cosmological parameters,
or the luminosity distance to a redshift bin. The connection between $\theta$ and
the data is specified in the likelihoods $P(\dat_j|\pp,\ldots)$ which in general
compare the measured magnitude to the theoretical value, with the theoretical value
depending on the $\theta$, in other words $P(\dat_j|\pp,\Ia)=\LL_{\Ia,j}(\theta)$,
and so on. The full formula then is 
\bea
P(\pp|\dat) &\propto& \sum_{b_k,\Sigma_k} P(\pp) P(b) P(\Sigma) \times \nonumber\\
&&\prod_{j=1}^N \Large\{ P(\dat_j|\pp,\Ia) P(\Ia)_j  \nonumber\\ 
&&+ P(\dat_j|\pp,b_\Ibc,\Sigma_\Ibc,\Ibc)P(\Ibc)_j \\
&&+ P(\dat_j|\pp,b_\II,\Sigma_\II,\II)P(\II)_j  \nonumber\\
&&+ P(\dat_j|\pp,b_X,\Sigma_X,X) \nonumber\\
&& \times (1-P(\Ia)_j-P(\Ibc)_j-P(\II)_j)  \Large\} \nonumber
\eea

If on the other hand we do not trust the absolute values of the $\chi^2$
then we can either add a bias to safeguard against a systematic shift in the absolute
probabilities, or allow for a global $P_X$ that an object is no supernova at all.
For this we always normalise the supernova probabilities to unity,
$P_j = \pi_j/\sum_i \pi_i$, and use the likelihood
\bea
P(\pp|\dat) &\propto& \sum_{b_k,\Sigma_k,P_X} P(\pp) P(b) P(\Sigma) P(P_X)\times \nonumber\\
&&\prod_{j=1}^N \Large\{ \Large[P(\dat_j|\pp,\Ia) P(\Ia)_j  \nonumber\\ 
&&+ P(\dat_j|\pp,b_\Ibc,\Sigma_\Ibc,\Ibc)P(\Ibc)_j \\
&&+ P(\dat_j|\pp,b_\II,\Sigma_\II,\II)P(\II)_j\Large] (1-P_X)   \nonumber\\
&&+ P(\dat_j|\pp,b_X,\Sigma_X,X) P_X  \Large\} \nonumber
\eea
It is probably a good idea to always run an analysis with additional
safeguards like this, and preferably a free global bias in the Ia
probability, in parallel to the ``real'' analysis in case something
goes very wrong. The global bias $\Delta$ might be added as
\bea
P(\pp|\dat) &\propto& P(\pp) \!\!\!\!\!\!\!\! \sum_{b_k,\Sigma_k,\Delta_i,P_X} \!\!\!\!\!\!\!\! P(P_X)
\!\!\!\!\!\!\!\! \prod_{k\in\{\Ibc,\II,X\}}\!\!\!\!\!\!\!\!
P(b_k) P(\Sigma_k)\prod_{i=1}^2  P(\Delta_i) \times \nonumber\\
&&\prod_{j=1}^N \Big\{ \Big[P(\dat_j|\pp,\Ia) \left[P(\Ia)_j-\Delta_1-\Delta_2\right]
  \nonumber\\ 
&&+ P(\dat_j|\pp,b_\Ibc,\Sigma_\Ibc,\Ibc)\left[P(\Ibc)_j+\Delta_1\right]
\label{fpost} \\
&&+ P(\dat_j|\pp,b_\II,\Sigma_\II,\II)\left[P(\II)_j+\Delta_2\right]
\Big] (1-P_X)   \nonumber\\
&&+ P(\dat_j|\pp,b_X,\Sigma_X,X) P_X  \Big\}, \nonumber
\eea

Especially the bias $\Delta_2$ of the Ia vs II probability is useful
to catch problems due to dust-reddening which can lead to a confusion
between these two classes \cite{dustpap}.

While estimating a dozen additional parameters is not really a problem
statistically if we have several thousand data points, it can become
a rather difficult numerical problem which justifies some work in
itself. We are using a Markov-chain monte carlo code with several
simulated annealing cycles to find the global maximum of the
posterior, which seems to work reasonably well but could certainly
be improved upon.

We notice that in addition to a measurement of the model parameters
$\theta$ from the Ia supernovae, we also get estimates of the distributions
of the other populations. In principle we could feed this information
back into the analysis.
Even though the prospect of being able to use the full information from all
data points is very tempting, we may not win much from doing so. We would
expect that the type-Ia supernovae are special in having a very small
dispersion in the absolute magnitudes. As such, they carry a lot more
information than a population with a larger dispersion. In terms
of our toy-example where $\sigma_A=0.1$ and $\sigma_B=2$  we need 
$(\sigma_B/\sigma_A)^2=400$ times more population $B$ points to achieve
the same reduction in the error. Unless we are lucky and discover
another population with a very small dispersion (or a way to make it so),
we expect that the majority of the information will always come from
the SNIa.

\section{Conclusions}

We present a generalised Bayesian analysis formalism called BEAMS (Bayesian 
Estimation Applied to Multiple Species) that provides a
robust method of parameter estimation from a contaminated data set
when an estimate of the probability of contamination is provided. The
archetypal example we have in mind is cosmological parameter
estimation from Type Ia supernovae (SNIa) lightcurves which will
inevitably be contaminated by other types of supernovae. In this case
lightcurve template analysis provides a probability of being a SNIa
versus the other types.

We have shown that BEAMS 
allows for significantly improved estimation when compared
to other estimation methods, which introduce biases and
errors to the resulting best-fit parameters.

BEAMS applies to the case where the probability, $P_i$, of the $i$-th point belonging to each of 
the underlying distributions is known. Where the data points are independent,
repeated marginalisation and application of Bayes' theorem
yields a posterior probability distribution that consists of a weighted sum of
the underlying likelihoods with these probabilities.
Although the general, correlated, case where the likelihood does not factor
into a product of independent contributions is simple to write down, it
contains a sum over $2^N$ terms (for 2 populations and $N$ data points).
This exponential scaling makes it unsuitable for application to real
data where $N$ is easily of the order of a few thousand. This case
will require further work.

We have studied in some detail the simple case
of estimating the luminosity distance in a single
redshift bin from one population consisting of SNIa candidates and
another of non SNIa candidates. In addition to an optimal
estimate of the luminosity distance, by including the
free shift $b$ and width $\Sigma$ of the wide Gaussian
distribution as variables in the MCMC estimation method, the BEAMS
method also allows one to gain insight into the underlying
distributions of the contaminants themselves, which is not possible
using standard techniques. Provided that the model for at least
one class of data are known, this method can be expanded to more
distributions, each with their own shift $b_i$ and width
$\Sigma_i$.

BEAMS was tested against other methods, such as using only a
spectroscopically confirmed data set in a $\chi^2$ analysis; using
only data points with probabilities higher than a certain cut off
value, and weighting a $\chi^2$ value by some function of the
probability. The Bayesian method performs significantly better than
the other methods, and provides optimal use of the data available. In
the SNe Ia case, the Bayesian framework provides an excellent platform
for optimising future surveys, which is specifically valuable given
the high costs involved in the spectroscopic confirmation of
photometric SNe candidates.

A Bayesian analysis is optimal if the underlying model is the true
model. Unfortunately in reality we rarely know what awaits us, and it
is therefore a good idea to add some extra freedom to the analysis,
guided by our experience. In this way BEAMS can also be applied when 
the population probability is not known precisely. In this
case a global uncertainty is added to the known probability
distributions, which can be estimated from the data. In the case of
the SNe Ia, one can include a global probability $p$ that the
supernovae belong to either group, and then marginalise over it,
allowing the data to not only estimate the most likely value for
$p$ but also to separate the data into the two classes. This
global approach can protect against outliers when the accuracy of the
type-probability is not known precisely. It is one of the strengths of Bayesian
approaches that they allow one to add quite general deviations from perfect data,
which are then automatically eliminated from the final result, and to
compute the posterior probability that such surprises were present.

A robust method of application of BEAMS to data from future
supernova surveys is proposed to estimate the properties of the
contaminant distributions from the data, and to obtain values for the
desired parameters.
Although we have illustrated and developed the BEAMS algorithm here
with explicit references to a cosmological application, it is far
more general. It can be easily applied to other fields, from photometric
redshifts to other astronomical data analyses and even to other fields
like e.g. biology. Since it is Bayesian in nature, it can very easily
be tailored to the specific needs of a subject, through simple and
straightforward calculations.

\begin{acknowledgements}

We thank Rob Crittenden and Bob Nichol for useful discussions and Joshua Frieman, Alex Kim and Pilar Ruiz-Lapuente for very useful comments on the manuscript. MK acknowledges support from the Swiss NSF, RH acknowledges support from NASSP.

\end{acknowledgements}

\bibliography{art}
\bibliographystyle{apsrev1}
\end{document}